# FERMION GREEN FUNCTIONS IN NON-ABELIAN GAUGE THEORIES IN FOUR DIMENSIONS

G. Triantaphyllou[1]

Department of Physics, University of Toronto
Toronto, ON M5S 1A7, Canada

**Abstract**

Previous results on fermion chirality-flipping four-point functions are extended to non-abelian gauge theories. The problem is purely non-perturbative, and the analytical formalism used is based on the Schwinger-Dyson hierarchy. This is truncated, and the resulting equations are solved numerically by relaxation techniques. Taking the large-$N$ limit in $SU(N)$ theories simplifies the problem substantially, allowing the formulation of conjectures on the behavior of $n$-point and chirality-flipping fermion Green functions for general $n$.

In this work we study the dynamical generation of fermion chirality flipping four-point functions within the framework of non-abelian gauge theories. These are generated in a purely non-perturbative manner, so in our analysis we have to use an adequate formalism like the one based on the Schwinger-Dyson (SD) hierarchy. The resulting equations are analytically intractable and very challenging even numerically. However, a certain limit in the theory allows a drastic simplification.

Our interest is focused on two objectives. One is to estimate the critical value of the gauge coupling, below which the four-point functions are equal to zero. We would like to study this in a case where four-fermion condensates develop on scales higher than the scale of two-fermion-condensate formation. This would make it consistent to treat the four-fermion condensate problem independently of the mass generation problem. Our other goal is to study the momentum dependence of the fermion four-point function. If such an object could generate a fermion or a gauge-boson mass, then two or 4 lines of the four-point function would be closed off into a loop or into another four-point function. It would therefore be interesting to study the details of the momentum dependence, such as the relative size of the four-point function when different pairs of momenta are large.

In the present study we use a one gauge-boson exchange approximation, where one gauge boson can attach to any pair of the four legs. The present work constitutes a clear progress with respect to our previous study [2], since, apart from considering a general non-abelian group, it includes a treatment of non-linearities, it does not neglect terms proportional to external momenta while at the same time exploring the full available momentum space, and in a certain limit allows us to draw conclusions on the behavior of general $n$-point functions.

---

[1]Talk given at the 17th annual MRST meeting in Rochester (May 1995), based on work in collaboration with B. Holdom [1].

We focus our attention on fermion operators that are purely chirality changing of the form $\bar{\psi}_L \Gamma \psi_R \bar{\psi}_L \Gamma' \psi_R + \text{h.c.}$, where $\psi_L \equiv \frac{(1-\gamma_5)}{2}\psi$ and $\psi_R \equiv \frac{(1+\gamma_5)}{2}\psi$. We constrain our study to just one fermion flavor. The four independent operators which have this property and respect parity are[2]

$$\begin{aligned}
\frac{1}{2}(\bar{\psi}\psi\bar{\psi}\psi + \bar{\psi}\gamma^5\psi\bar{\psi}\gamma^5\psi) &= \bar{\psi}_L\psi_R\bar{\psi}_L\psi_R + \text{h.c.} \\
\bar{\psi}\sigma^{\mu\nu}\psi\bar{\psi}\sigma^{\mu\nu}\psi &= \bar{\psi}_L\sigma^{\mu\nu}\psi_R\bar{\psi}_L\sigma^{\mu\nu}\psi_R + \text{h.c.} \\
\frac{1}{2}(\bar{\psi}\lambda^a\psi\bar{\psi}\lambda^a\psi + \bar{\psi}\lambda^a\gamma^5\psi\bar{\psi}\lambda^a\gamma^5\psi) &= \bar{\psi}_L\lambda^a\psi_R\bar{\psi}_L\lambda^a\psi_R + \text{h.c.} \\
\bar{\psi}\lambda^a\sigma^{\mu\nu}\psi\bar{\psi}\lambda^a\sigma^{\mu\nu}\psi &= \bar{\psi}_L\lambda^a\sigma^{\mu\nu}\psi_R\bar{\psi}_L\lambda^a\sigma^{\mu\nu}\psi_R + \text{h.c.}, \quad (1)
\end{aligned}$$

where $\lambda^a$ are the generators of the non-abelian gauge group we wish to study. From now on we will denote all operators of the form $\bar{\psi}\Gamma\psi\bar{\psi}\Gamma'\psi$, where $\Gamma, \Gamma'$ are matrices with possibly non-trivial spinor and color structure, by $\Gamma \otimes \Gamma'$.

The vacuum expectation values of these four operators receive exclusively non-perturbative contributions, and their non-zero values would break dynamically a chiral $U(1)_A$, were the latter not broken by instantons.

The SD formalism relevant to these four-fermion operators was discussed in Ref. [2] and led to an equation shown diagrammatically in Fig. 1. It results from a truncation of the SD hierarchy achieved by approximating the four-fermion+photon vertex appearing in the equation by a four-fermion vertex and a photon attached via a bare vertex on one of its legs. Such an approximation neglects, among others, non-linearities that would become important in the infra-red. We will return to them later. The equation has been symmetrized to include diagrams with gluons connecting all possible pairs of fermions, so the right-hand side is multiplied by a factor of 1/2.

We are considering a four-point function associated with the Green function $\langle 0 | T\{\bar{\psi}_\alpha \psi_\beta \bar{\psi}_\rho \psi_\tau\} | 0 \rangle$. In momentum space, the four-point function receiving exclusively non-perturbative contributions is

$$\begin{aligned}
\mathcal{O}_{\alpha\beta\rho\tau} &= \mathcal{O}_{S+P}(I_{\alpha\beta} \otimes I_{\rho\tau} + \gamma^5_{\alpha\beta} \otimes \gamma^5_{\rho\tau}) + \mathcal{O}_T \sigma^{\mu\nu}_{\alpha\beta} \otimes \sigma^{\mu\nu}_{\rho\tau} \\
&+ \mathcal{O}^{color}_{S+P}(\lambda^a I_{\alpha\beta} \otimes \lambda^a I_{\rho\tau} + \lambda^a \gamma^5_{\alpha\beta} \otimes \lambda^a \gamma^5_{\rho\tau}) + \mathcal{O}^{color}_T \lambda^a \sigma^{\mu\nu}_{\alpha\beta} \otimes \lambda^a \sigma^{\mu\nu}_{\rho\tau} \quad (2)
\end{aligned}$$

where the four scalar functions $\mathcal{O}_{S+P}$, $\mathcal{O}_T$, $\mathcal{O}^{color}_{S+P}$ and $\mathcal{O}^{color}_T$ depend on 6 variables, which are all the independent and Lorenz-invariant combinations of the external 4-momenta $p_1, ..., p_4$, corresponding to the fermions with spinor indices $\alpha, \beta, \rho, \tau$ respectively. We wish to develop the SD equations for these scalar functions.

The functional integral operators $\Gamma^i$ appearing below are defined exactly as in our previous work [2]. We work in the Landau gauge which is popular in studies of SD equations; the gauge boson propagator reads $\frac{D^{\mu\nu}}{k^2} \equiv \frac{1}{k^2}\left(\delta^{\mu\nu} - \frac{k^\mu k^\nu}{k^2}\right)$.

---

[2] We choose to work in Euclidean space, so there is no difference between upper and lower Lorentz indices.

By combining the results of the abelian case [2] with considerations pertaining on the color structure in a $SU(N)$ theory, we have

$$\mathcal{O}_{S+P} = \frac{3(N^2-1)}{2N}(\overline{\Gamma^A}+\overline{\Gamma^B})[\mathcal{O}_{S+P}] + 6(\overline{\Gamma^C}+\overline{\Gamma^D}-\overline{\Gamma^E}-\overline{\Gamma^F})[\mathcal{O}_T^{color}]$$

$$\mathcal{O}_T = -\frac{N^2-1}{2N}(\overline{\Gamma^A}+\overline{\Gamma^B})[\mathcal{O}_T] + 2(\overline{\Gamma^C}+\overline{\Gamma^D}+\overline{\Gamma^E}+\overline{\Gamma^F})[\mathcal{O}_T^{color}] +$$

$$\frac{1}{2}(\overline{\Gamma^C}+\overline{\Gamma^D}-\overline{\Gamma^E}-\overline{\Gamma^F})[\mathcal{O}_{S+P}^{color}]$$

$$\mathcal{O}_{S+P}^{color} = -\frac{3}{2N}(\overline{\Gamma^A}+\overline{\Gamma^B})[\mathcal{O}_{S+P}^{color}] + \frac{3(N^2-1)}{2N^2}(\overline{\Gamma^C}+\overline{\Gamma^D}-\overline{\Gamma^E}-\overline{\Gamma^F})[\mathcal{O}_T] +$$

$$6\left(-\frac{N^2+2}{2N}(\overline{\Gamma^C}+\overline{\Gamma^D})+\frac{1}{N}(\overline{\Gamma^E}+\overline{\Gamma^F})\right)[\mathcal{O}_T^{color}]$$

$$\mathcal{O}_T^{color} = \left(\frac{1}{2N}(\overline{\Gamma^A}+\overline{\Gamma^B})+\left(-\frac{N^2+2}{N}(\overline{\Gamma^C}+\overline{\Gamma^D})-\frac{2}{N}(\overline{\Gamma^E}+\overline{\Gamma^F})\right)\right)[\mathcal{O}_T^{color}] +$$

$$\frac{N^2-1}{2N^2}(\overline{\Gamma^C}+\overline{\Gamma^D}+\overline{\Gamma^E}+\overline{\Gamma^F})[\mathcal{O}_T] + \frac{N^2-1}{8N^2}(\overline{\Gamma^C}+\overline{\Gamma^D}-\overline{\Gamma^E}-\overline{\Gamma^F})[\mathcal{O}_{S+P}] +$$

$$\frac{1}{2}\left(-\frac{N^2+2}{2N}(\overline{\Gamma^C}+\overline{\Gamma^D})+\frac{1}{N}(\overline{\Gamma^E}+\overline{\Gamma^F})\right)[\mathcal{O}_{S+P}^{color}], \qquad (3)$$

where $\overline{\Gamma^i}[\mathcal{O}] \equiv \Gamma^i(\mathcal{O}^i)$ and $\mathcal{O}^i$ are the form factors with loop-dependent arguments corresponding to diagrams $i = A, ..., F$ [2].

We are faced with a system of four coupled 4-dimensional integral equations involving functions of 6 variables. As it stands, the problem is analytically intractable, and even a numerical solution proves to be beyond our present means.

By taking the large-$N$ limit, Eq.3 reduces to

$$\begin{pmatrix} \mathcal{O}_{S+P} \\ \mathcal{O}_T \\ \mathcal{O}_{S+P}^{color} \\ \mathcal{O}_T^{color} \end{pmatrix} = N \begin{pmatrix} \frac{3}{2}(\overline{\Gamma^A}+\overline{\Gamma^B}) & 0 & 0 & 0 \\ 0 & -\frac{1}{2}(\overline{\Gamma^A}+\overline{\Gamma^B}) & 0 & 0 \\ 0 & 0 & 0 & -3(\overline{\Gamma^C}+\overline{\Gamma^D}) \\ 0 & 0 & -\frac{1}{4}(\overline{\Gamma^C}+\overline{\Gamma^D}) & -(\overline{\Gamma^C}+\overline{\Gamma^D}) \end{pmatrix} \begin{pmatrix} [\mathcal{O}_{S+P}] \\ [\mathcal{O}_T] \\ [\mathcal{O}_{S+P}^{color}] \\ [\mathcal{O}_T^{color}] \end{pmatrix}$$

(4)

It is clear that in the large-$N$ limit, the two form factors $\mathcal{O}_{S+P,T}$ decouple from the other two $\mathcal{O}_{S+P,T}^{color}$. As regards the first two, it is obvious that only $\mathcal{O}_{S+P}$ is going be non-zero, since $\mathcal{O}_T$ receives a negative contribution from the dominant diagrams $A$ and $B$. It is also apparent that $\mathcal{O}_{S+P}^{color}$ can be non-zero only if $\mathcal{O}_T^{color}$ is non-zero. Having a non-zero $\mathcal{O}_T^{color}$ seems highly unlikely, however, since it receives a negative contribution from the dominant diagrams $C$ and $D$ involving itself. Even if it were able to be non-zero, it would require a critical coupling much larger than the one required for $\mathcal{O}_{S+P}$ and it should therefore not present a problem.

The form that this system of coupled integral equations takes in the large-$N$ limit allows us to argue for the omission of contributions coming from operators proportional to external momenta. With regards to the operators with structure $p_i^\mu \sigma^{\mu\nu} \otimes p_j^\lambda \sigma^{\nu\lambda}$ and $p_i^\mu p_j^\nu \sigma^{\mu\nu} \otimes p_k^\lambda p_l^\tau \sigma^{\lambda\tau}$, which have so far been neglected, we can see that the contributions they receive from the dominant diagrams having the same functions in their vertices are negative, so we do not expect the corresponding four-point functions to be non-zero. Similar arguments make us neglect operators with structure $\mathcal{O}_1 \equiv p_i^\mu p_j^\nu \sigma^{\mu\nu} \otimes \mathbf{1}$ and $\mathcal{O}_2 \equiv \mathbf{1} \otimes p_i^\mu p_j^\nu \sigma^{\mu\nu}$ (and trivial color structure).

We now write down the integral equation for $\mathcal{O}_{S+P}$ according to the previous discussion. We have

$$\mathcal{O}_{S+P} = \frac{N\alpha}{16\pi^3} \int d^4k \gamma^\mu \frac{1}{\slashed{k}(\slashed{k}-\slashed{q})} \gamma^\nu \left( \frac{{}^k\mathcal{O}_{S+P}^A D_1^{\mu\nu}}{(k-p_1)^2} + \frac{{}^k\mathcal{O}_{S+P}^B D_4^{\mu\nu}}{(k-p_4)^2} \right) \qquad (5)$$

where $q \equiv p_1 + p_2$, $D_i^{\mu\nu} \equiv \delta^{\mu\nu} - \frac{(k-p_i)^\mu (k-p_i)^\nu}{(k-p_i)^2}$ and the left-superscript of ${}^k O_{S+P}^{A,B}$ is a reminder that these functions depend on the loop momentum.

We go in a reference frame where $\vec{p_1} = -\vec{p_2}$, and the form of the kernel of the integral equation indicates that $\mathcal{O}_{S+P}$ is a function of five variables, i.e. $\mathcal{O}_{S+P} = \mathcal{O}_{S+P}(p_1^0, |\vec{p_1}|, p_4^0, |\vec{p_4}|, q^0)$. We also have ${}^k\mathcal{O}_{S+P}^A = {}^k\mathcal{O}_{S+P}^A(k^0, |k|, p_4^0, |\vec{p_4}|, q^0)$ and ${}^k\mathcal{O}_{S+P}^B = {}^k\mathcal{O}_{S+P}^B(p_1^0, |\vec{p_1}|, k^0, |k|, q^0)$.

The structure of the equation indicates that a special solution to the above integral equation is factorizable, i.e. $\mathcal{O}_{S+P} \sim \tilde{B}(p_1^0, |\vec{p_1}|, q^0) \tilde{B}(p_4^0, |\vec{p_4}|, q^0)$. Physically this is expected, since the momenta flowing in diagram $A$ are independent from the momenta flowing in diagram $B$. Setting $p_1^0 = p_4^0 \equiv p^0$ and $|\vec{p_1}| = |\vec{p_4}| \equiv |p|$, we have

$$\tilde{B}(p^0, |p|; q^0) = \frac{N\alpha}{8\pi^3} \int d^4k \tilde{B}(k^0, |k|; q^0) \gamma^\mu \frac{1}{\slashed{k}(\slashed{k}-\slashed{q})} \gamma^\nu \frac{D_1^{\mu\nu}}{(k-p_1)^2} \qquad (6)$$

where we have separated $q^0$ from the other two variables with a semi-column, since it does not mix with loop momenta and can thus be treated as a parameter.

Note that for $q^0 = 0$ this reduces to the linearized SD equation for the fermion two-point function. For non-zero $q^0$, the problem reduces to the one of a three-point function.

We have so far neglected non-linearities coming in our problem. These come from diagrams containing an odd number of four-point functions and a gluon in a way that makes them important only in the IR regime. They therefore provide our equation with an effective IR cut-off. Their non-linear nature gives our solution a natural scale. These diagrams are not easily resummable, so we multiply the kernel of our equation by the function $A(q^0) \equiv \frac{k^2}{k^2 + \Lambda^2 \tilde{B}^2(0,0;q^0)}$, where $\Lambda$ is the effective IR cut-off in our theory provided by the non-linearities, in an effort to mimic their effects, and we normalize $\tilde{B}(0,0;0) = 1$. Our full solution will thus be given by $\mathcal{O}_{S+P} = \left(\frac{r}{\Lambda}\right)^2 \tilde{B}(p_1^0, |\vec{p_i}|, q^0) \tilde{B}(p_4^0, |\vec{p_4}|, q^0)$, where $r$ is a constant larger than unity because of the numerical factors appearing in the loops by which non-linearities are introduced in the problem.

Moreover, we change our variables from $p^0$, $k^0$, $|p|$, and $|k|$, to $p = \sqrt{p^{0\,2} + |p|^2}$, $k = \sqrt{k^{0\,2} + |k|^2}$, $\phi = \arctan(|p|/p^0)$ and $\tilde{\phi} = \arctan(|k|/k^0)$, so $\tilde{B}(p^0, |p|; q^0) \equiv B(p, \phi; q^0)$. We then solve for the function $pB$ in order to increase the accuracy of the numerical solution that follows, since we expect $B$ to decrease with increasing $p$. After some Dirac algebra, and omitting terms with structure $p_i^\mu p_j^\nu \sigma^{\mu\nu} \otimes \mathbf{1}$ according to our previous discussion, the equation becomes

$$pB(p, \phi; q^0) = \frac{3N\alpha}{4\pi^2} \int dk d\tilde{\phi} \frac{kB(k, \tilde{\phi}; q^0) k \sin\tilde{\phi}}{\left((q^0 - k\cos\tilde{\phi})^2 + k^2 \sin^2\tilde{\phi}\right) 2\sin\phi} \times$$

$$\ln\left(\frac{(k\sin\tilde{\phi} + p\sin\phi)^2 + (k\cos\tilde{\phi} - p\cos\phi)^2}{(k\sin\tilde{\phi} - p\sin\phi)^2 + (k\cos\tilde{\phi} - p\cos\phi)^2}\right) \times$$

$$\left(1 - \frac{q^0 k \cos\tilde{\phi}}{k^2}\right) \times \frac{k^2}{k^2 + \Lambda^2 B^2(0,0;q^0)}, \quad (7)$$

where $0 \leq \phi, \tilde{\phi} \leq \pi$.

The form of the integral equation allows us to use the same discretization lattice for the arguments of the function $B$ inside and outside the integral. We choose the number of lattice points in each dimension to be equal to 40. Our results do not change by varying this number. The integral equation is solved via a relaxation method. We first insert an initial configuration for our function, and then iterate the equation until it is satisfied to a reasonable accuracy.

We start by setting $q^0 = 0$. This gives us a critical coupling $N\alpha_c = 2.7 \pm 0.3$. This coupling is somewhat larger than the one required for a non-zero two-point function, which, for large-$N$, is given by the relation $N\alpha_c = \frac{2\pi}{3} \approx 2.1$. This discrepancy is expected to disappear in the limit $\Lambda_{UV} \longrightarrow \infty$. We then keep this coupling fixed, and for different values of $q^0$ we compute the corresponding values of $B(0,0;q^0)$. In Fig. 2 we plot the function $\frac{p}{\Lambda} B(p, \phi; 0)$.

We see that $\frac{p}{\Lambda} B(p, \phi; 0)$ exhibits the $\cos(\gamma \log(p/\Lambda))$ behavior ( with $\gamma$ a coupling-dependent constant and $\Lambda$ the IR cut-off of the equation) that is well known in two-point function studies, and that it is independent from $\phi$, as expected. We do not trust the behavior of the function for momenta smaller than $\Lambda$, since we expect our formalism to break down at that region. As $q^0$ increases and gets larger than $\Lambda B(0,0;q^0)$, it plays the role the non-linearities played before and acts like an effective IR cut-off, so the shape of $B$ is the same as the one of the two-point function having $q^0$ as its IR cut-off. The dependence on $\phi$ remains very weak. This behavior was expected for $q^0 = 0$, and it persists for larger $q^0$. Moreover, $B$ falls rapidly with increasing $q^0$, so the non-linearities become increasingly irrelevant in this regime.

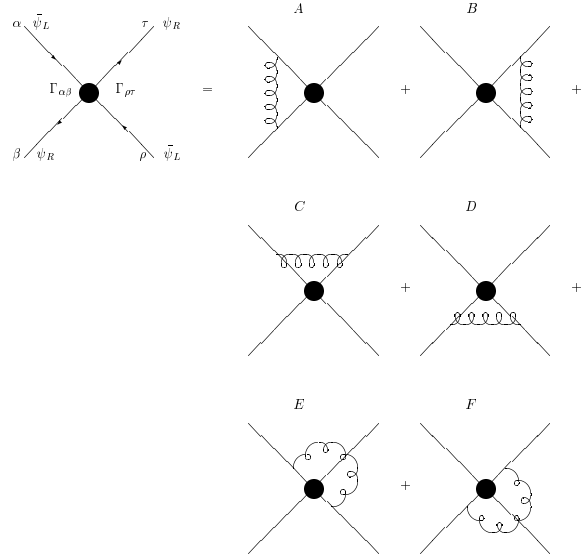

Figure 1: The schematic form of the SD equation. We have labeled the four fermions by their spinor indices. We label the diagrams by the capital letters $A, ..., F$, and we omit the factor of $1/2$ multiplying the right-hand side.

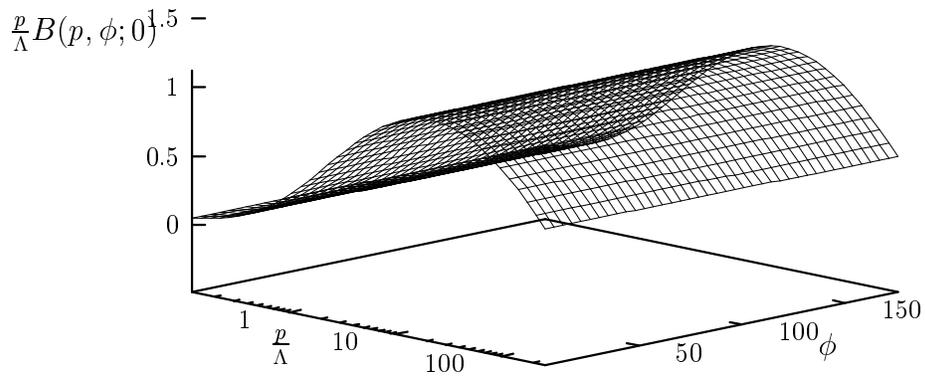

Figure 2: The function $\frac{p}{\Lambda}B(p, \phi; 0)$

The behavior found is approximately $B(0,0;q^0) \approx \left(1 - \frac{q^0}{1.16\Lambda}\right)^{\frac{1}{2}}$. It is obvious that, for a given critical coupling, $q^0$ cannot exceed $\Lambda$ by much before the whole solution collapses to zero.

The form that the integral equation takes for large $N$ suggests a possible generalization of the above results. In particular, the diagrams that dominate are the ones in which a gluon is attached between two fermion fields with spinor indices contracted with each other, since color flows through a closed loop. Therefore, one might expect only condensates of the form $\frac{1}{2} < (\bar{\psi}\psi...\bar{\psi}\psi + \bar{\psi}\gamma^5\psi...\bar{\psi}\gamma^5\psi) > = < \bar{\psi}_L\psi_R...\bar{\psi}_L\psi_R + \text{h.c.} >$, where the ellipsis stands for $n - 4$ fermion fields paired with each other as in the first and last pair, to get non-zero expectation values at the same critical value of the gauge coupling as the 2- and four-point functions.

Moreover, we expect the corresponding scalar $n$-point function $O^n_{S+P}$ to factorize and to be given by $O^n_{S+P} = \Lambda_c^p B_1...B_k$, where $\Lambda_c$ is the characteristic energy of the theory, and $p = 4 - 3k$ with $k = n/2$. In analogy with the previous discussion, we have defined the functions $B_i \equiv B(p_i, \phi_i; q_i), i =, 1..., k$, where, as before, $p_i$ is the four-momentum of one of the fermions in each pair and $q_i$ is the total four-momentum of the pair. In order for each of the functions $B_i$ to satisfy the same final integral equation as before, however, we have to assume that the three-momenta of the fermions in each pair are equal and opposite to each other, so that $\vec{q}_i = 0$. The qualitative results should not change substantially, nevertheless, if one makes the weaker assumption that we can go in an optimum reference frame where no fermion has a three-momentum which is much larger than the three-momentum of the fermion it is contracted with.

To conclude, in this work we have treated a truncated Schwinger-Dyson equation for the four-point function of fermions interacting via a non-abelian gauge interaction in the Landau gauge. We considered chirality-changing four-point functions which receive non-perturbative contributions exclusively. The large-$N$ limit of $SU(N)$ renders our problem analogous to a three-point function one. Our numerical results indicate that the equations exhibit a critical behavior, and that the critical coupling appears to be roughly equal to the one required for the formation of two-fermion condensates. The beauty of the large-$N$ limit is that it also allows us to guess the critical and momentum behavior of fermion chirality-flipping $n$-point functions for higher $n$.